\documentclass[prl,twocolumn]{revtex4}

\usepackage{graphicx}
\usepackage{textcomp}
\usepackage{float}
\usepackage{epstopdf}
\usepackage{color}
\usepackage{amsmath}
\usepackage{amssymb}
\usepackage{amsfonts}

\newcommand{\vari}{\textcolor[rgb]{1.00,0.00,0.00}}

\newcommand{\RR}{\right}
\newcommand{\LL}{\left}
\newcommand{\m}{\mathrm}

\begin{document}

\title{Multimode circuit optomechanics near the quantum limit}

\author{Francesco Massel$^1$}
\author{Sung Un Cho$^1$}
\author{Juha-Matti Pirkkalainen$^1$}
\author{Pertti J. Hakonen$^1$}
\author{Tero T. Heikkil\"a$^1$}
\author{Mika A. Sillanp\"a\"a$^1$}
\thanks{Present Address: Department of Applied Physics, Aalto University School of Science, P.O. Box 15100, FI-00076 Aalto, Finland}

\affiliation{$^1$Low Temperature Laboratory, Aalto University, P.O. Box 15100, FI-00076 AALTO, Finland}

%\affiliation{$^1$Low Temperature Laboratory, Aalto University, P.O. Box 15100, FI-00076 AALTO, Finland}

%\date{\today}

\begin{abstract}
The coupling of distinct systems underlies nearly all physical phenomena and their applications. A basic instance is that of interacting harmonic oscillators, which gives rise to, for example, the phonon eigenmodes in a crystal lattice. Particularly important are the interactions in hybrid quantum systems consisting of different kinds of degrees of freedom. These assemblies can combine the benefits of each in future quantum technologies. Here, we investigate a hybrid optomechanical system having three degrees of freedom, consisting of a microwave cavity and two micromechanical beams with closely spaced frequencies around 32 MHz and no direct interaction. We record the first evidence of tripartite optomechanical mixing, implying that the eigenmodes are combinations of one photonic and two phononic modes. We identify an asymmetric dark mode having a long lifetime. Simultaneously, we operate the nearly macroscopic mechanical modes close to the motional quantum ground state, down to 1.8 thermal quanta, achieved by back-action cooling. These results constitute an important advance towards engineering entangled motional states.
\end{abstract}

\maketitle

%\fnsymbol{3}
%\footnotetext{Present address: Department of Applied Physics, Aalto University School of Science, P.O. Box 15100, FI-00076 Aalto, Finland } 

%\vspace{0.3cm}
%\pagebreak

One of the present goals in physics is the explanation of
macroscopic phenomena as emerging from the quantum-mechanical laws
governing nature on a microscopic scale. This understanding is also important for future quantum information applications \cite{DeMille,PainterRoute,SembaQBspin,KorppiPRL2011}, since many of the most promising platforms base on nearly macroscopic degrees of freedom. For macroscopic mechanical objects, their potential quantum behavior \cite{OConnell:2010br} has been actively investigated with resonators interacting with an electromagnetic cavity mode \cite{KippenbergReview}. In particular, freezing of the mechanical Brownian motion \cite{Metzger:2004ei,Rocheleau:2010jd}  down to below one quantum of energy has recently been observed \cite{Teufel2011b,AspelmeyerCool11}.

Another result of radiation-pressure interaction is the mixing of the normal modes of vibration into linear combinations of the uncoupled phonon and photon modes. This has been verified with optomechanical Fabry-Perot cavity \cite{Groeblacher:2009eh}, and by its on-chip microwave analogy \cite{Teufel2011b} using an aluminum membrane. All this work is paving the way towards engineering non-classical motional \cite{OConnell:2010br} and hybrid quantum states \cite{DeMille,LaHaye2009,SembaQBspin,KorppiPRL2011} for basic tests of quantum theory \cite{Legget2002,MartinisBell}, as well as applications in foreseeable future.

%operating at dilution refrigerator temperatures down to a few tens of mK.

Corresponding phenomena in systems comprising of more than two active degrees of freedom  \cite{KippenbergMulti,MarquardtMulti} have received less attention. In the optomechanical crystal setup, coupling of mechanical vibrations via radiation pressure interaction has been demonstrated with the zipper cavity \cite{PainterMix2010}, however, with some direct interaction between the beams. In the microwave regime, such measurements have remained outstanding. In this work, we take a step further, and examine a multimode system where two micromechanical beams, having resonant frequencies $\omega_1/(2\pi) = 32.1$ MHz and $\omega_2/(2\pi) = 32.5$ MHz, are each coupled to a microwave on-chip cavity. We obtain the first evidence of hybridization of all the three degrees of freedom. This was made possible by operating, as opposed to the optical setup, in the good-cavity limit, that is, both $\omega_1$ and $\omega_2$ were much larger than the cavity decay rate $\gamma_c$. Simultaneously, the mechanical modes are found at low occupation numbers $n_m$ near the quantum ground state, down to $n_m = 1.8 \pm 0.5$. These are the lowest occupation recorded with nanowire resonators to date, and they occur via the sideband cooling, starting from operation at dilution refrigerator temperatures at a few tens of mK.

\begin{figure}[h]
\centering
\includegraphics[width=0.95\linewidth]{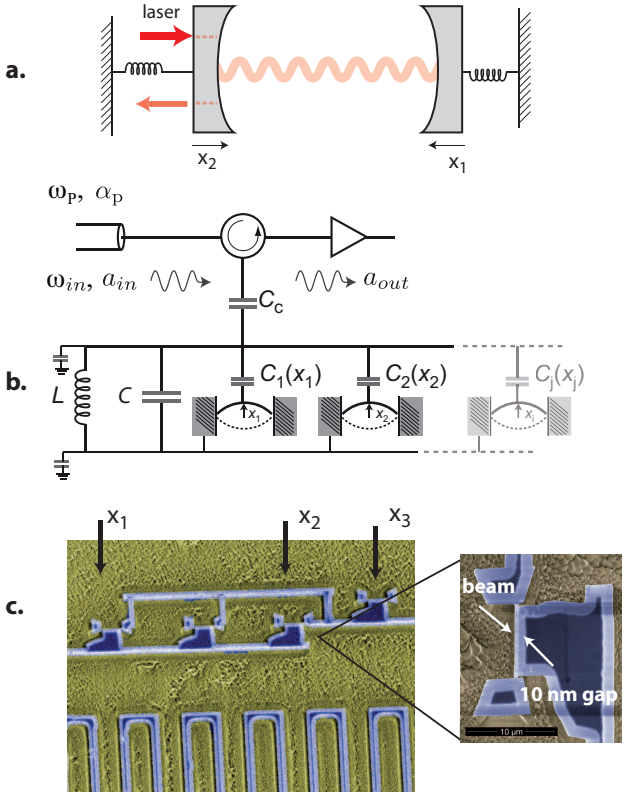}
\caption{\textbf{Cavity-controlled coupling of micromechanical resonators}. \textbf{a}, Optomechanical system with both mirrors movable has three degrees of freedom. \textbf{b}, analogy of (a) with a microwave-regime cavity represented by the lumped element inductors and capacitors and arbitrary number $j=1...N$ of mechanical resonators. The experiments are performed by shining a strong pump microwave tone (frequency $\omega_P$), and possibly a weak probe tone (frequency $\omega_{\m{in}}$) in a dilution refrigerator down to 25 mK temperature. \textbf{c}, images of the all-aluminum superconducting sample, showing four fabrication jigs for a beam. Total of three beams ($x_1 ... x_3 $) were operating. Beams 1 and 2 had large electromechanical couplings $g_1/2\pi = 1.8$ MHz/nm, $g_2/2\pi = 2.0$ MHz/nm, whereas the third beam had an order of magnitude smaller coupling. }\label{fig1}
\end{figure}

In the optical domain \cite{PainterMix2010}, a three-mode optomechanical system can be pictured as a Fabry-Perot cavity with both massive mirrors mounted via springs, see Fig.~1a. In the microwave version, (Fig.~1b), the mirrors are replaced by movable capacitors formed by the mechanical resonators. The electromechanical coupling arises when each mechanical resonator (labelled with $j$), expressed as time-varying capacitors $C_j$, independently modulates the total capacitance, and hence, the cavity frequency $\omega_{c}$. This is described by the coupling energies $g_j = (\omega_c / 2 C) \partial C_j/ \partial x_j$.

In the actual device (see micrograph in Fig.~1c), the mechanical resonators are spatially separated by about 100 microns, and we hence expect the direct interaction between the beams via the substrate material to be negligible. We drive the cavity strongly by a microwave pump signal applied at the frequency $\omega_{P}$ near the cavity frequency. The pump induces a large cavity field with the number of pump photons $n_P \gg 1$. This allows for a linearized description of the electromechanical interaction and results in substantially enhanced effective couplings $G_j = g_j x_{0j} \sqrt{n_P}$.
\section*{Results}
\vspace{-0.3cm}
\textbf{Cavity with several mechanical modes} The general Hamiltonian for $N$ mechanical resonators with frequencies $\omega_j$, individually coupled to a common cavity mode, written in a frame rotating with the pump, becomes
\begin{align}
\label{eq:h2}
H = - \hbar \Delta a^\dag a + \hbar \sum_{j=1}^N \omega_j b_j^\dag b_j  - \hbar  \LL(a^\dag + a \RR) \sum_{j=1}^N G_j \LL(b_j^\dag + b_j \RR)
\end{align}
Here, $ a$ is the annihilation operator for the cavity mode, $ b_j$ are those of the mechanical resonators, and $x_{0j} = \sqrt{\hbar/2 m_j \omega_j}$ is the zero-point amplitude. If one assumes roughly similar mechanical resonators, viz.~ $\omega_j \sim \omega_m$ and $G_j \sim G$, the cavity becomes nearly resonant to all of them at the red-sideband detuned pump condition $\Delta = -\omega_m$. 

Coupling of the mechanical resonators via the cavity "bus" can be anticipated to be significant if the mechanical spectra begin to overlap when their width increased by the radiation pressure, $\gamma_{\m{eff}} =\gamma_m + \gamma_{\m{opt}}$, grows comparable to the mechanical frequency spacing. Here, $\gamma_{\m{opt}}= 4 G^2/\gamma_c$. The equations of motion following from Eq.~(\ref{eq:h2}) allow one to verify the above assumption. We will in the following focus on two resonators, $N=2$. The result of such calculation, which is detailed in the Supplementary Information, is that they experience an effective coupling with energy $G_{12} = G_1 G_2/\gamma_c$. 

In the limit of strong coupling, $\sqrt{(G_1^2+G_2^2)/2} \gg \gamma_c/4$, the system is best described as a combination of two "bright" modes with linewidths $\sim \gamma_c/2$, and of a "dark" mode having a long lifetime $\sim \gamma_m$. In the latter, the two mechanical resonators oscillate out-of-phase and synchronize into a common frequency $(\omega_1+\omega_2)/2$. In any case, the mode structure can be written with the normal coordinates for mode $n$, for position quadrature according to $X^{(n)} = D_c^{(n)} q_c + \sum_{j=1}^N D_j^{(n)} q_j$, and similarly for the momentum quadrature $P^{(n)}$. 
 
In the experiment, we use doubly clamped flexural beams which are made by the use of ion beam milling of aluminum \cite{Sulkko:2010ih}, having an ultranarrow $\sim 10$ nm vacuum slit to the opposite end of the cavity for maximizing the coupling. The cavity is $\lambda/2$ floating microstrip similar to Ref.~ \cite{MechAmpPaper}, resonating at $\omega_c/(2 \pi) = 6.98$ GHz. The total cavity linewidth $\gamma_{c} = \gamma_E + \gamma_I \simeq (2 \pi)
\times 6.2$ MHz is a sum of the internal damping $\gamma_I/(2\pi) = 1.4$ MHz, and the external damping $\gamma_E/(2\pi) = 4.8$ MHz. At at the highest powers discussed, however, we obtain a decreased $\gamma_I/(2\pi) = 0.9$ MHz, typical of dielectric loss mechanism.

There are a total of three beams as shown in Fig.~1c. Two of the beams have a large zero-point coupling of $g_1 x_{01} /2\pi = 39$ Hz, and $g_2 x_{02} /2\pi = 44$ Hz. The frequencies of beams 1 and 2 were relatively close to each other, $\omega_2 - \omega_1 \sim (2 \pi) \, 450$ kHz, such that it is straightforward to obtain an effective coupling $G_{12}$ of the same order. The third beam had an order of magnitude smaller coupling, and hence we neglect it in the full calculation, setting $N=2$ in Eq.~(\ref{eq:h2}) in what follows.

\begin{figure*}
\centering
\includegraphics[width=0.95\linewidth]{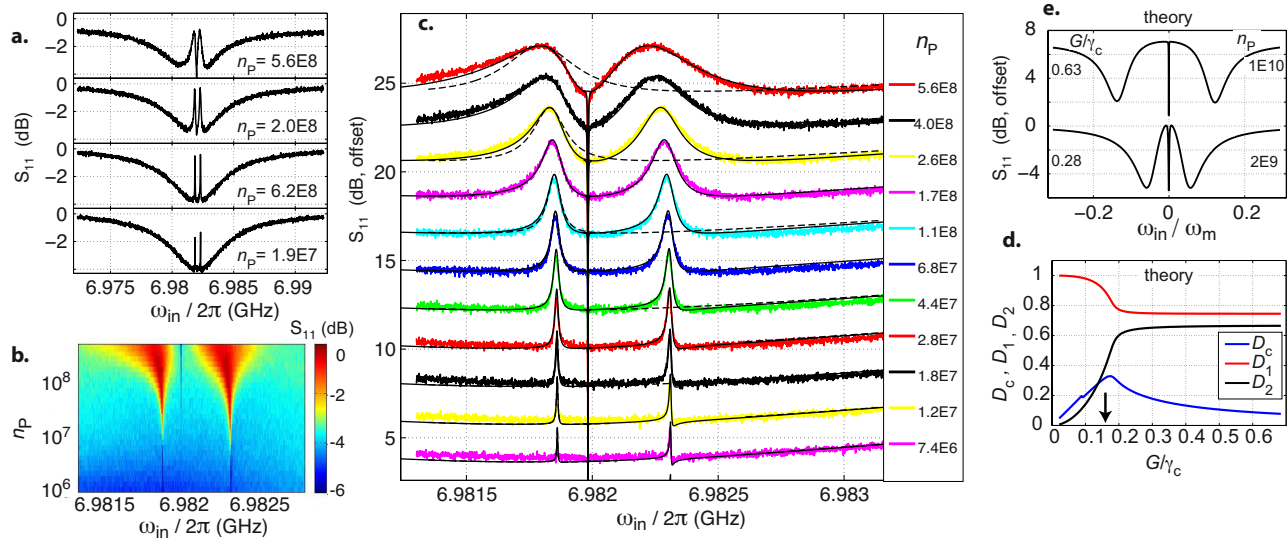}
\caption{\textbf{Hybridization of microwave photons and two radio-frequency phonons}. \textbf{a}, Measured reflection coefficient of probe microwave while a strong pump is applied at the red sideband, $\Delta \simeq - \omega_1$, while increasing pump amplitude, from bottom to top. \textbf{b}, \textbf{c}, zoomed-in view of (a). An excellent agreement with the theory is obtained with reflection calculation using a complete model based on Eq.~(\ref{eq:h2}), for two beams 1 and 2 (solid lines). The dashed lines show a fit with an independent-resonator model for beam 1. The narrow feature at 6.982 GHz is due to beam 3. The curves are displaced vertically by 3 dB for clarity. \textbf{d}, normal modes of oscillations expressed as linear combinations of photon and two phonons for the dark mode. The parameters are as in  \textbf{c}. The arrow denotes the highest experimental value. \textbf{e}, Theory plot for higher $n_P$; the dark mode is expected to become narrower and more and more phonon-like. Here, $\Delta = -(\omega_1 + \omega_2)/2$. For simplicity, we take $g^{(2)}_j = 0$ in \textbf{d} and \textbf{e}.}\label{fig2}
\end{figure*}

The measurements are conducted in a cryogenic temperature below the superconducting transition temperature of the Al film ($\sim 1$ K) in order to reduce losses. We used a dilution refrigerator setup as in Ref.~\cite{MechAmpPaper} down to 25 mK, which allows for a low mechanical mode temperature in equilibrium. Apart from this, cryogenic temperatures or superconductivity are not necessary \cite{Weig2011}.

\vspace{0.4cm}
\textbf{Experimental data}
The modes can be probed by adding to the experimental setup another, the probe, tone \cite{Groeblacher:2009eh,Teufel:2011ha,MechAmpPaper} with frequency $\omega_{\m{in}}$. The maximum pump power we can reach is $4 \,\mu$W, inducing a coherent $n_P \sim 6 \times 10^8$ photon occupancy in the cavity, and effective coupling $G_{12} =(2\pi)\times 150$ kHz (with $G_1 = 0.93$ MHz, $G_2 = 1.0$ MHz). In Fig.~2a, we display changes of the cavity absorption by increasing the pump power. The two peaks at the bottom of the cavity dip are due to the beams 1 and 2. They broaden with increasing $n_P$, finally leaving behind a narrow dip between them. The overall lineshape is clearly not a sum of two Lorentzian peaks.

In Fig.~2c we show a zoom-in view of the two peaks, together with theoretical predictions. One immediately sees that an attempt to simulate either peak by a single mechanical resonator coupled to cavity ($N=1$ in the analysis), thereby neglecting the cavity-mediated coupling, fails to explain the lineshape at large $n_P \gtrsim 10^8$. The full simulation with both beams 1 and 2 included, however, produces a remarkable agreement to the experiment. In order to create these curves, we further take into account a shift of the mechanical frequencies due to second-order interaction \cite{Rocheleau:2010jd}, given as $-\frac{1}{2} g^{(2)}_j n_P$, where $g^{(2)}_j = x_{0j} ^2 \omega_c /2C \LL( \partial^2 C_j/\partial x^2_j \RR) \sim (2\pi) \, 3 \times 10^{-4}$ Hz is the second-order coupling coefficient. The third beam, visible in Fig.~2c as a narrow feature just below 6.982 GHz, was later fitted by running a single-resonator calculation. This is justified because its mixing is negligible owing to weak coupling.

The narrow dip between the peaks manifests the onset of the dark mode, where the cavity participation approaches zero with growing $G_j$, as occurs also in other tripartite systems \cite{Tripartite}. As usual for weakly decaying states, the dark modes can be useful for storage of quantum information \cite{EITrmp,PainterMix2010}. The other two dips (best discerned in the theoretical plots at higher $G_j$, see Fig.~2e) are the bright modes, and they retain fully mixed character of all three degrees of freedom. In Fig.~2d we display the prediction for the mode expansion coefficients with respect to coupling. Asymmetry in the role of the mechanical modes is sensitive to parameters. With the maximum coupling in the experiment ($G_j/\gamma_c \simeq 0.15$), the normal modes are well mixed: the dark mode can be expressed as $\LL[D_c, D_1, D_2 \RR] \sim \LL[0.3, 0.3, 0.9 \RR]$.

\begin{figure}[h]
\label{figdark}
\includegraphics[width=0.95\linewidth]{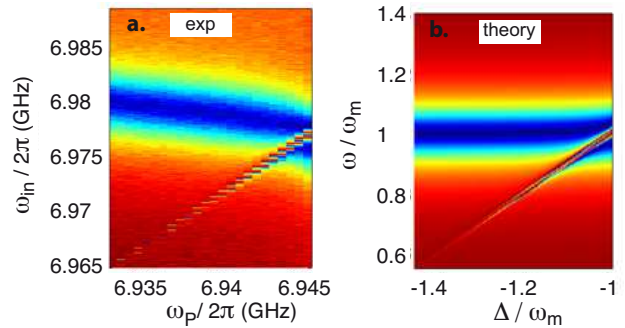}
\caption{\textbf{Pump detuning measurements.} \textbf{a}, Measured spectrum of the tripartite cavity electromechanical system when pumped at a fixed input power such that at resonance, $\Delta \sim \LL( -\omega_{1}, -\omega_{2} \RR)$, $n_P \simeq  5.6\times 10^8$. The pump frequency was slowly swept from left to right. \textbf{b}, theory prediction.}
\end{figure}
\begin{figure*}[ht]
\centering
\includegraphics[width=0.9\linewidth]{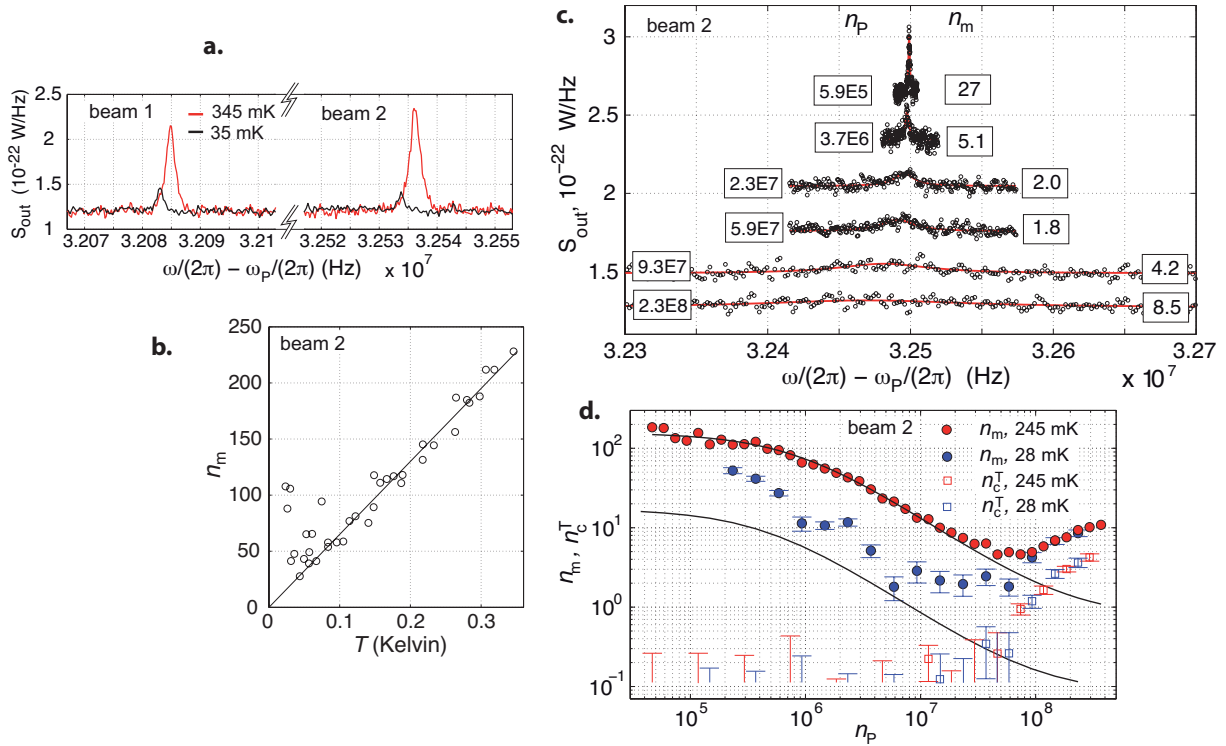}
\caption{\textbf{Sideband cooling of motion down to 1.8 thermal quanta}. \textbf{a}, cavity output spectra representing the thermal motion peaks of beams 1 and 2 measured at 345 mK and 35 mK, with $n_P = 4 \times 10^5$.  \textbf{b}, calibration of the detection according to temperature dependence of the mechanical energy $ n_m \hbar \omega_m = k_B T$, given by the area of the peak (data for beam 2). \textbf{c}, output spectra of beam 2 at pump occupation $n_P$ increasing from top to bottom. The curves are shifted vertically by 0.3 y-axis units for clarity. The corresponding mechanical occupations are written on the right. \textbf{d}, progress of back-action cooling (beam 2) as a function of pump occupation. The mechanical mode temperature is given by circles, and the cavity temperature by squares. Red: high temperature, 245 mK; blue, base temperature (28 mK). The ideal theoretical behavior are plotted by the solid lines. Unless the data point for $n_c^T$, including its error bars, was of positive sign, only the absolute value of the error bar is plotted. }\label{figcool}
\end{figure*}

One can also vary the pump detuning, thereby effectively detuning the cavity and the mechanical resonators \cite{Groeblacher:2009eh,Teufel:2011ha}. An anticrossing of the cavity and mechanical modes is seen in Fig.~3a up to detuning $\Delta \sim -\omega_m$, above which the system exhibits parametric oscillations. The pertinent simulation, Fig.~3b, portrays the main features of the measurement, except bending of the cavity frequency towards lower value in an upward sweep of $\Delta$. We expect such possibly hysteretic behavior of the cavity frequency to be due to second-order effects beyond the present linear model.

\vspace{0.4cm}
\textbf{Sideband cooling}
We now turn the discussion into showing how the tripartite system resides nearly in a pure quantum state during the mode mixing experiments. The mechanical resonators can be sideband-cooled with the radiation pressure back-action \cite{Regal:2008di,Kippenberg2009,Hertzberg:2010,Rocheleau:2010jd} under the effect of the pump which also creates the mode mixing. Although the ground state of one mode was recently achieved this way \cite{Teufel2011b,AspelmeyerCool11}, cooling in multimode or nanowire systems is less advanced.

The incoherent output spectrum can be due to either a finite phonon number $n_m$, or non-equilibrium population number $n_c^T$ of the cavity. The former manifests as a Lorentzian centered at the upper sideband, as in Fig.~4a, with the linewidth $\gamma_{\m{eff}} = \gamma_m + 4 G^2/\gamma_c$, whereas the latter gives a small emission at broader bandwidth. The analytical theory \cite{Rocheleau:2010jd} and experimental details are discussed in Supplementary Information. 

At low input power, when the cavity back-action damping is small and $n_c^T \sim 0$, phonon number is set by thermal equipartition $n_{mj} \hbar \omega_{j} = k_B T$. This is best observed by the area of the mechanical peak in the output spectrum, as in Fig.~4a, which should be linear in cryostat temperature. The linearity holds down to about 150 mK temperatures (Fig.~4b), below which we observe intermittent heating which is sensitive to parameters.

The resulting cooling of beam 2 is displayed in Fig.~4d. At a relatively high temperature of 245 mK, the data points follow theory well up to input powers corresponding to $n_P \sim 5 \times 10^7$. At the minimum cryostat temperature of 28 mK, the mode is not thermalized in a manner which depends irregularly on power. We obtain a minimum phonon number $n_{m2} \sim 1.8$, where the mechanical mode spends one third of the time in the quantum ground state. The other mechanical mode, beam 1, cooled simultaneously down to 2.5 quanta, possibly compromised by the slightly smaller coupling. The cooling is quite clearly bottlenecked by heating, by the pump microwave, of the bath to which the mechanical mode is coupled. For the optimum data points, the starting temperature for cooling has raised to 150 mK, and above this it grows roughly quadratically with $n_P$. 

\section*{Discussion}
\vspace{-0.3cm}
The coupling of micromechanical resonators mediated via microwave photons in an on-chip cavity is a basic demonstration of the control of a multimode mechanical system near the quantum limit. The setup provides a flexible platform for creation and studies of nonclassical motional states entangled over the chip \cite{OptoEntang}, or over macroscopic distances \cite{AspelmeyerPRL07,ligo}. The setup is easily extended to embody nearly arbitrarily many mechanical resonators, hence allowing for designing an electromechanical metamaterial with microwave-tunable properties. For future quantum technology applications even at elevated temperatures \cite{Weig2011}, data may be stored \cite{Painter2011} in weakly decaying dark states identified in the present work.

\section*{Methods}
\small
\vspace{-0.3cm}
\textbf{Device fabrication}

The lithography to define the meandering cavity and jigs for the beams consisted of a single layer of electron-beam exposure, followed by evaporation of 150 nm of aluminum on a fused silica substrate. Since the cavity is of planar structure, it has no potentially problematic cross-overs which tend to have weak spots in superconducting properties and hence could heat up the cavity. We thus expect the cavity to portray a high critical current limited only by the intrinsic properties of the strip. With a 2 $\mu$m wide strip, we expect the critical current to be several mA. With the maximum circulating currents in cavity (Fig.~2), about 0.5 mA of peak value, there was no sign of nonlinearity in the cavity response.

The mechanical resonators were defined by Focused Ion Beam (FIB) cutting, as in Ref.~\cite{Sulkko:2010ih}. We used low gallium ion currents of 1.5 pA which gives the nominal beam width of about 7 nm. In order to maximize sputtering yield with minimal gallium contamination, we used a single cutting pass mode. Otherwise, the cut beams tend to collapse to the gate. With the current recipe, fabrication of down to 10 nm gap widths over 5-10 microns distance can be done with about 50 \% yield.

\vspace{0.3cm}

\textbf{Theory of cavity-coupled resonators}

The Hamiltonian for $N$ mechanical resonators each coupled to one cavity mode via the radiation pressure interaction is
\begin{equation}
\label{eq:h1}
H = \hbar \omega_c a^\dag a + \hbar \sum_{j=1}^N \omega_j b_j^\dag b_j  - \hbar  a^\dag a \sum_{j=1}^N g_j x_{0j} \LL(b_j^\dag + b_j \RR) +H_P
\end{equation}
The pump with the Hamiltonian $H_P = 2 \hbar \sqrt{\frac{P_{P} \gamma_E}{\hbar \omega_c}} \LL( a^{\dagger} + a \RR) \cos {\omega_P t}$ induces a large cavity field with the number of pump photons $n_P \gg 1$. 

Under the strong driving, the cavity-mechanics interaction can be linearized individually for each beam, resulting in Eq.~(\ref{eq:h2}). Further details of the calculations are given in the Supplementary Information.

\vspace{0.3cm}
\textbf{Sideband cooling}

For one beam, the measured output spectrum divided by system gain, $S_{\m{out}}(\omega)$, carries information on the phonon number according to \cite{Rocheleau:2010jd}
\begin{align}
\label{eq:cool}
S_{\m{out}} = \frac{\gamma_E}{\gamma_c} n_c^T  + \frac{\gamma_E}{\gamma_c}\gamma_{\m{opt}}\frac{\gamma_{\m{eff}}}{\LL(\omega-\omega_m\RR)^2 + \gamma_{\m{eff}}^2/4} \LL( n_m - 2 n_c^T \RR)
\end{align}
which is a Lorentzian centered at the cavity frequency. A non-zero base level is due to possible broadband emission from the cavity, due to a thermal state with occupation $n_c^T$. We suppose that at modest pump (cooling) powers of utmost interest, coupling of the mechanical resonators remains insignificant, and thus we can apply Eq.~(\ref{eq:cool}) separately for each resonator. Some deviations can, however, occur at the highest powers with $n_P \sim 3 \times 10^8$.

In order to obtain the phonon number under strong back-action, we fit a Lorentzian to each peak as in Fig.~4c, obtaining independently the base level, amplitude and linewidth for every input power. These values are then compared to Eq.~(\ref{eq:cool}). This leaves, however, yet too many unknowns in order to obtain the occupations of the mechanics and cavity. This is basically because of uncertainties in the attenuations of both input and amplifier lines, which limit the accuracy of estimating $n_P$. We can obtain a calibration using the linear temperature dependence, and by fitting the dependence of $\gamma_{\m{eff}}$ on input power.

As opposed to for instance to Ref.~\cite{Rocheleau:2010jd}, the bottleneck for cooling is not heating of the cavity by strong pump, as seen in Fig.~4d, where the cavity temperature is increasing only little up to the optimum cooling powers corresponding to about 0.15 mA of peak current values in cavity. This can be due to the simplistic structure of the cavity.

The above analysis supposes the presence of nearly no non-equilibrium photons in the cavity at zero or low input powers. This situation can be investigated by observing a possible emission from about cavity linewidth under these conditions. A complication arises because the cavity linewidth is so large such that a small change is easily overwhelmed by modest standing waves in cabling. We avoided this issue by using temperature dependence of the cavity frequency, due to the kinetic inductance in the long microstrip. Thereby, a broad emission peak moving according to the cavity frequency would be easily distinguishable. We observed no such signal down to the level of $n_P^T \lesssim 0.1$ which justifies the above analysis.

\vspace{1cm}
\normalsize
\textbf{Acknowledgements} We would like to thank S. Paraoanu and Lorenz Lechner for useful discussions. This work was supported by the Academy of Finland, by the European Research Council (grants No. 240362-Heattronics and 240387-NEMSQED), and by the Vaisala Foundation.

\bibliographystyle{naturemag}
\bibliography{/Users/masillan/Documents/latex/bib_ms}

%\begin{thebibliography}{10}

%\end{thebibliography}

%\bibitem{suppl}See supplementary material at http://link.aps.org/supplemental/.

%\textbf{Supplementary Information} is linked to the online version.

%\textbf{Author Contributions} F.M. and T.H. developed the theory. J.-M.P., S.U.C. and M.A.S. contributed to design and fabrication of the samples, and cryogenic setup.  P.J.H. and M.A.S. conceived the work. 

%\textbf{Author Information} The authors declare no competing financial interests. Correspondence and requests for materials should be addressed to M.A.S. (Mika.Sillanpaa@aalto.fi).

\pagebreak
%\onecolumn

\begin{widetext}
%\begin{multicols}{3}

\Large{Multimode circuit optomechanics near the quantum limit: Supplementary information}

\normalsize

\vspace{0.5cm}
Francesco Massel$^1$,
Sung Un Cho$^1$,
Juha-Matti Pirkkalainen$^1$,
Pertti J. Hakonen$^1$,
Tero T. Heikkil\"a$^1$, and
Mika A. Sillanp\"a\"a$^1$

$^1$Low Temperature Laboratory, Aalto University, P.O. Box 15100, FI-00076 AALTO, Finland

\section{Response of coupled resonators}
\label{Two_b}
We detail here the theory of the response of a driven cavity coupled
to $N$ (with emphasis on the case $N=2$) mechanical resonators. In
particular, we show how it is possible to describe the system from two
complementary points of view. On one hand, the cavity-mechanics
interaction provides a dressing of the dynamics of the resonators,
leading to the definition of effective mechanical frequencies and
dampings. On the other hand, for strong coupling, we find that on top
of the normal mode splitting found for single resonators, the system
exhibits a dark mode which gets asymptotically decoupled from the
cavity, and whose linewidth tends to the bare linewidth of the
mechanical resonances.

We define the following symbols: The mechanical resonators have frequencies $\omega_j$ and linewidths $\gamma_j$. Their annihilation operators are denoted by $ b_j$ while $q_j = \frac{1}{\sqrt{2}}\LL( b_j^\dag+b_j\RR) $ and $p_j= \frac{i}{\sqrt{2}}\LL( b_j^\dag-b_j\RR)$ are the dimensionless  amplitude and momentum of the mechanical
oscillations. The mechanical zero-point amplitude is $x_{0j} = \sqrt{\hbar/2 m_j \omega_j}$. Correspondingly, the cavity has the frequency $\omega_c$ and linewidth $\gamma_c=\gamma_E+\gamma_I$, consisting of the external and internal dissipation, respectively. $a$ is the annihilation operator for the cavity. The pump microwave (applied at the frequency $\omega_{P}$ near the cavity frequency) has the detuning from the cavity $\Delta=\omega_P-\omega_c$, and the power $P_P$.

The Hamiltonian for $N$ mechanical resonators each coupled to one cavity mode via the radiation pressure interaction is
\begin{equation}
\label{eq:hS1}
H = \hbar \omega_c a^\dag a + \hbar \sum_{j=1}^N \omega_j b_j^\dag b_j  - \hbar  a^\dag a \sum_{j=1}^N g_j x_{0j} \LL(b_j^\dag + b_j \RR) +H_P
\end{equation}
The pump with the Hamiltonian $H_P = 2 \hbar \sqrt{\frac{P_{P} \gamma_E}{\hbar \omega_c}} \LL( a^{\dagger} + a \RR) \cos {\omega_P t}$ induces a large cavity field with the number of pump photons $n_P \gg 1$. The number of quanta the in cavity due to pump  is
\begin{equation} 
\label{nc}
n_P =  \frac{P_{P} \gamma_E}{\hbar \omega_c} \frac{1}{\Delta^2 + \LL(\frac{\gamma}{2}\RR)^2} 
\end{equation} 
where angular frequency units are used.

The equations of motion following from the Hamiltonian, Eq.~(1) in the main text, linearized about a steady-state, can be written as \cite{masselsuppl}
\begin{align}
 \label{eq:da}
  \dot{a}&=i \Delta a - \frac{\gamma_c}{2} a+
  i\sum_{j=1}^N G_j q_j +
  \sum_{i=I,E}\sqrt{\gamma_{i}} a^{i}_{\rm in}  \\
  \label{eq:dq1}
     \dot{q}_j&= \omega_j  p_j\\
\label{eq:dp1}
     \dot{p}_j&= - \omega_j  q_j - \gamma_{j} p_j +
     G_j \left(a^\dagger+a\right)  + \xi_j, 
\end{align}
where $a_{\rm in}^i$
are input fields to the cavity ($I$ and $E$ denoting internal and
external input fields, the previous describing internal dissipation
and the latter the coupling to the measurement setup), and $\xi_j$ are
fields driving the mechanical resonators.  
Moreover,
$G_j=\sqrt{n_P}g_j$ describes the cavity driving (by photon number
$n_P$) enhanced coupling between the cavity field and resonator $j$.  Fourier transforming and, after some algebra, assuming
a probe field $a_{\rm in}=a_{\rm in}^E$ and neglecting the noise terms
$\xi_j$ yields
\begin{align}
%  &\left({\omega_j}^2-\omega^2-i\gamma_j\omega\right) 
%    q_j =  G_j  \omega_i a^\dagger+a  \\
%&\left[-i\left(\omega+\Delta\right)+\gamma_c/2\right] a=i\sum_{j=1}^N 
%G_j q_j+\sqrt{\gamma_E} a_{\rm in} \\
&X \equiv a^\dagger+a=
          \tilde{\chi} \left(\omega \right) \sum_{j=1}^N G_j q_j
             -i   \sqrt{\gamma_E}\left[
                          \chi_{-\Delta}\left(\omega\right) a_{\rm in}+
                          \chi_{\Delta}\left(\omega\right)
                          a_{\rm in}^{\dagger}
                 \right]\\
 &\label{eq:1}q_j=\chi_j(\omega) G_j X
\end{align}
where the response functions are
\begin{equation}
  \label{eq:2}
  \chi_{\Delta}(\omega)=\frac{1}{\Delta-\omega-i\gamma_c/2}, \quad \chi_j(\omega)=\frac{\omega_j}{\omega_j^2-\omega^2-i \gamma_j \omega}
\end{equation}
and
\begin{equation}
  \label{eq:3}
  \tilde{\chi}(\omega)= \chi_{-\Delta}(\omega)-\chi_{\Delta}(\omega)=\frac{2\Delta}{(\omega+i\gamma_c/2)^2-\Delta^2}.
\end{equation}
Solving for $X$, we get
\begin{equation}
  \label{eq:4}
  X=\sqrt{\gamma_E} M(\omega)
  \left[\chi_{-\Delta}a_{\rm in}+\chi_{\Delta}a_{\rm in}^{\dagger} \right]
\end{equation}
with 
\begin{equation}
  \label{eq:5}
  M(\omega)=\frac{1}{1-\tilde{\chi}(\omega)\sum_{j=1}^N G_j^2 \chi_j(\omega)}.
\end{equation}
This field is coupled to the cavity output field by 
\begin{equation}
a_{\rm out}^{(\dagger)}=\sqrt{\gamma_E}a^{(\dagger)}-a_{\rm in}^{(\dagger)}.
\end{equation}
When the effect of $a_{\rm in}^\dagger$ on $a$ can be disregarded (see
below), we can write $a_{\rm out}=R(\omega) a_{\rm in}$, where
$R(\omega)=-i\gamma_E M(\omega) \chi_{-\Delta}(\omega) -1$ is the
reflection probability amplitude in the cavity. Figure 2 of the main
text shows its absolute value, corresponding to the measured
observable.

The effective dynamics of the system is qualitatively different in
different regimes. For typical structures $\gamma_c \gg \gamma_j$. In
this case, when $G_1^2+G_2^2 \ll (\gamma_c/4)^2$, the cavity response
$\tilde{\chi}(\omega)$ is approximatively independent of frequency
within the relevant frequencies describing the mechanics. In this case,
we can view the system as two mechanical resonators effectively
coupled by the cavity (Sec.~\ref{sec:lowcoupling}). On the other hand,
for large $G_j$ (sec. IB), the normal modes of the total system split. There are
two bright modes strongly coupled to the cavity, with resonance
frequencies $\omega_{b} \approx \Delta \pm \sqrt{\sum_j G_j^2}$ and
linewidths $\sim \gamma_c/2$, and one ``dark'' mode at around
$(\omega_1+\omega_2)/2$ and with a linewidth $\sim \gamma_j$. The
latter is almost decoupled from the cavity, with the coupling decreasing
with an increasing $G_j$. 

\subsection{Weak-coupling regime}
\label{sec:lowcoupling}

% which can be expressed as
% \begin{equation}
%   \label{eq:14}
%   M(\omega)=\frac{\left({\omega_1}^2-\omega^2-i\gamma_1\omega\right)
%                               \left({\omega_2}^2-\omega^2-i\gamma_2\omega\right)}
%                              {\left({\omega_1}^2-\omega^2-i\gamma_1\omega\right)
%                               \left({\omega_2}^2-\omega^2-i\gamma_2\omega\right)-
%                               \tilde{\chi}(\omega)(\sqrt{2} G_1\omega_1+\sqrt{2} G_2\omega_2)}
% \end{equation}
In the following, we concentrate on the case of red-detuned driving,
$\Delta \approx -\omega_j$ in the sideband resolved case $\gamma_c
\ll \omega_j$, and study the response at frequencies $\omega \approx
-\Delta$. This allows us to approximate
\begin{equation}
  \chi_j(\omega) \approx \frac{1}{2(\omega_j-\omega) -i\gamma_j}, \quad \tilde{\chi} \approx \frac{1}{\omega+\Delta+i\gamma_c/2}.
\label{eq:apprrespfuncs}
\end{equation}
The same approximation yields $\chi_\Delta(\omega) \ll
\chi_{-\Delta}(\omega)$ and we may hence disregard the input field
$a_{\rm in}^\dagger$ at these frequencies.

Let us restrict to the case of two mechanical resonators. Then the
response function $M(\omega)$ may be written in the form
\begin{align}
    M(\omega)=& \frac{\left(\omega_1-\omega-i\gamma_1/2\right)
                              \left(\omega_2-\omega-i\gamma_2/2\right)}
                             {\left(\omega_{1}-\omega-i\gamma_{1}/2\right)
                              \left(\omega_{2}-\omega-i\gamma_{2}/2\right)-
                          \tilde{\chi}(\omega)\left[ G_1^2\left(\omega_2-\omega-i
                          \gamma_2/2\right)+ G_2^2\left(\omega_1-\omega-i
                          \gamma_1/2\right)\right]/2} \nonumber\\
                     \equiv & \frac{\left(\omega_1+\omega-i\gamma_1/2\right)
                              \left(\omega_2+\omega-i\gamma_2/2\right)}
                             {\left(\omega_{1}^c+\omega-i\gamma_{1}^c/2\right)
                              \left(\omega_{2}^c+\omega-i\gamma_{2}^c/2\right)}
                             \label{eq:6b}
\end{align}
%  X=&\frac{M(\omega)}{\sqrt{2}}\chi_{-\Delta}(\omega) a_{\rm in} \\
%  q_i=&\frac{\sqrt{2} G_i} {({\omega_i}+\omega-i\gamma_i/2)} X
%\end{align}
We return to the definition of $\omega_j^{c}$ and $\gamma_j^{c}$
below, but now we rewrite the response in a more compact form
\begin{align}
  \label{eq:7}
  &X= \frac{\chi_{1}^c}{\chi_1} \frac{\chi_{2}^c}{\chi_2} \chi_{-\Delta} \sqrt{\gamma_E} a_{\rm in} \\\label{eq:q1}
  &q_1= G_1 \frac{\chi_{2}^c}{\chi_2}\chi_{1}^c \chi_{-\Delta} \sqrt{\gamma_E} a_{\rm in} \\\label{eq:q2}
  &q_2= G_2 \frac{\chi_{1}^c}{\chi_1}\chi_{2}^c \chi_{-\Delta}  \sqrt{\gamma_E} a_{\rm in}, 
\end{align}
where $\chi_j^c=\chi_j(\omega_j \mapsto \omega_j^c)$. 
This form is readily generalized for more than two mechanical resonators.
% \begin{figure}%[!ht]
%    \includegraphics[width=0.9\textwidth]{out1.eps}
%    \includegraphics[width=0.9\textwidth]{out2.eps}
%    \includegraphics[width=0.9\textwidth]{out3.eps}
%    \caption{Output spectrum ($|a_{out}|=|a-\sqrt{\gamma_E}a|$) for
%      increasing effective coupling $\sqrt{2} G$ ($G$ in the figures) }
%  \label{fig:a_out}
% \end{figure}
We now calculate $\omega_j^c$ and $\gamma_j^c$. For the sake of
compactness, we include the broadening to the frequencies, allowing
them to assume complex values
\begin{align}
  \label{eq:9}
   \omega_j\equiv\omega_j-i\gamma_j/2.
\end{align}
The frequencies $\omega_1^c$,  $\omega_2^c$ appearing in the
denominator of Eq.~\eqref{eq:6b} can be expressed as 
% \begin{equation}
%   \label{eq:10}
%   \omega^2 - \tilde{\chi}(\omega)/2 \left(\sqrt{2} G_1^2 \omega_2
%     +\sqrt{2} G_2^2 \omega_1 \right)-
%     \left[\omega_1-\tilde{\chi}(\omega)/2 \sqrt{2} G_1^2 +
%             \omega_2-\tilde{\chi}(\omega)/2 \sqrt{2} G_2^2\right]\omega
%     +\omega_1 \omega_2 =0.
% \end{equation}
% As we did in the amplification paper, we define
% \begin{align}
%   \label{eq:11}
%   \omega_1^e=\omega_1-\frac{\tilde{\chi}(\omega)}{2} \sqrt{2} G_1^2 \\
%   \omega_2^e=\omega_2-\frac{\tilde{\chi}(\omega)}{2} \sqrt{2} G_2^2 .
% \end{align}
% These frequencies represent the effective complex frequencies of the
% mechanical resonators, neglecting the cavity-induced coupling between
% the two resonators. 
% Solving eq. \eqref{eq:11}, we obtain
\begin{equation}
  \label{eq:12}
  \omega_{1,2}^c=\frac{1}{2}\left[\omega_1^e+\omega_2^e \pm 
                 \sqrt{\left(\omega_1^e-\omega_2^e\right)^2+\tilde{\chi}(\omega)^2 G_1^2 G_2^2}\right]
\end{equation}
where 
\begin{align}
  \omega_j^e=\omega_j-\tilde{\chi}(\omega) G_j^2/2 \label{eq:opticalspring}
\end{align}
are the effective frequencies (and
dampings) of the mechanical resonators induced by the presence of the
cavity, in the absence of the other mechanical resonator. The
frequencies (and dampings) $\omega_j^c$ are thus the dressed
mechanical frequencies in the presence of the cavity-mediated coupling
between the mechanical resonators. Equation \eqref{eq:12} describes an
avoided crossing of the two mechanical resonators coupled by the
energy
\begin{equation}
G_{12} \equiv \frac{1}{2}G_1 G_2 i\tilde{\chi}(\omega) \approx \frac{G_1 G_2}{\gamma_c}.
\label{eq:gmech}
\end{equation}
The latter form is valid in the limit $\omega_j^c+\Delta \ll
\gamma_c$, where we can approximate $\tilde{\chi} \approx
-2i/\gamma_c$. For strong coupling or large initial frequency
difference, the determination of $\omega_{1,2}^c$ requires the
solution of Eq.~\eqref{eq:12} with the replacement
$\tilde{\chi}(\omega) \to\tilde{\chi}(\omega_{1,2}^c)$. The solutions
of such a calculation are shown in Fig.~\ref{fig:cfreqs}, plotting the
resulting frequency shifts and changes in the damping.

When $G_{12}\ll \omega_1^e-\omega_2^e$, we can disregard the
coupling between the mechanical resonators, and the response of the
cavity is a simple product of the individual responses. In the
experiments, we reach $G_{12} =(2\pi)\times 150$ kHz (with $G_1 = 0.93$ MHz, $G_2 = 1.0$ MHz), which is
roughly one third of the frequency difference between the mechanical
resonators, and allows for the coupling effect to show up in the
shape of the response curves. This is shown in
Fig.~\ref{fig:comparison}, which compares the coupled response to the
case of individual resonators.

Although the dressed response is affected by the coupling, the
frequency shifts are not easily observed from the cavity response as
they tend to be overwhelmed by the growing width $\sim {\rm
  Im}(\omega_j^c)$. However, the frequency shifts can be seen in the
mechanical response $q(\omega)/a_{\rm in}$, plotted in
Fig.~\ref{fig:q_out}.
% \begin{figure}%[!ht]
%    \includegraphics[width=0.9\textwidth]{weff1.eps}
%    \includegraphics[width=0.9\textwidth]{weff2.eps}
%    \includegraphics[width=0.9\textwidth]{weff3.eps}
%    \caption{  $\operatorname{Re} \omega_{1,2}^c$ for  increasing effective coupling
%      $\sqrt{2} G$. In the last plot the (nearly) constant effective
%      frequency correspond to the dark-mode frequency, while the other
%      corresponds to the bright mode. }
%  \label{fig:a_out2}
% \end{figure}

 \begin{figure}%[!ht]
    \includegraphics[width=0.45\textwidth]{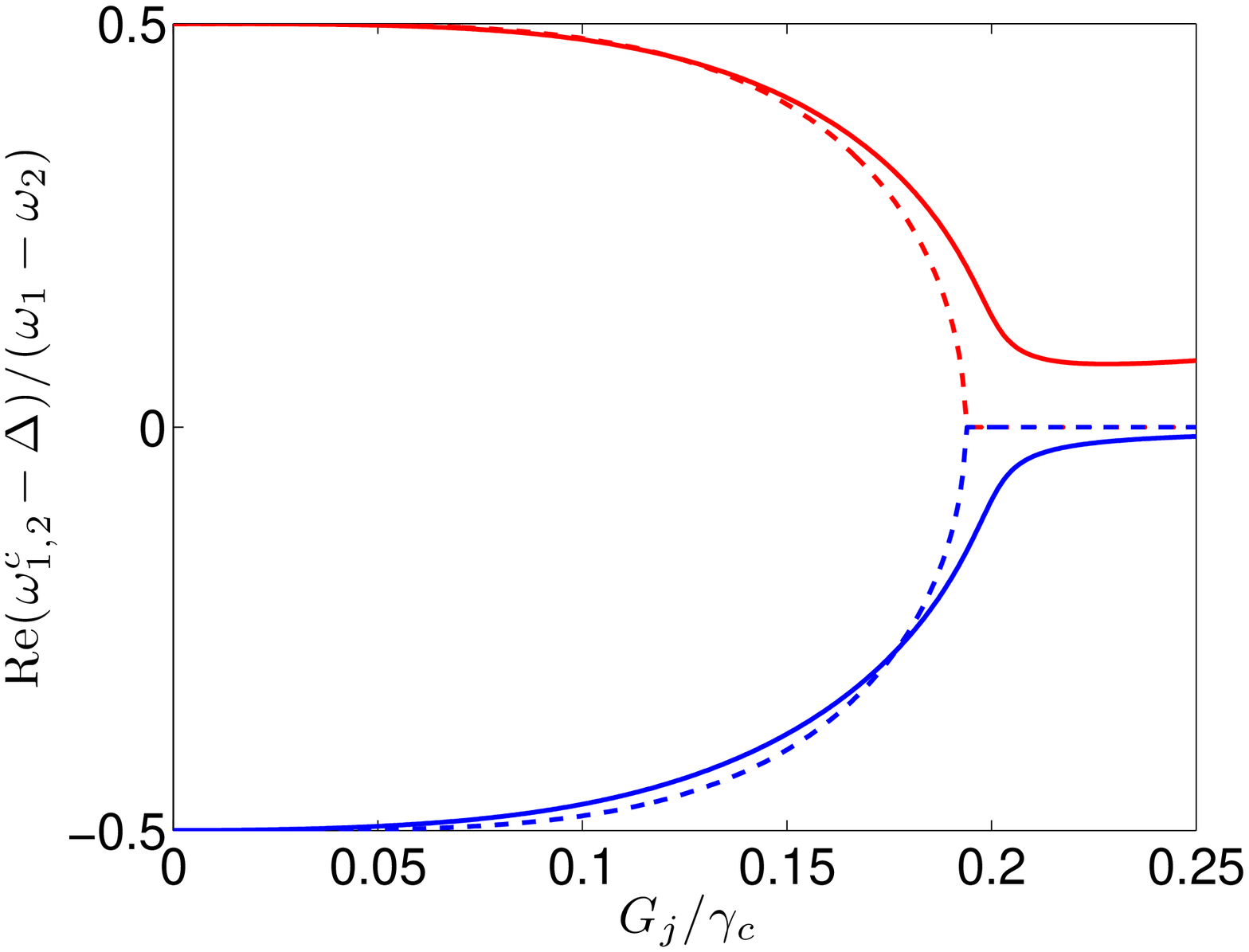}
    \includegraphics[width=0.45\textwidth]{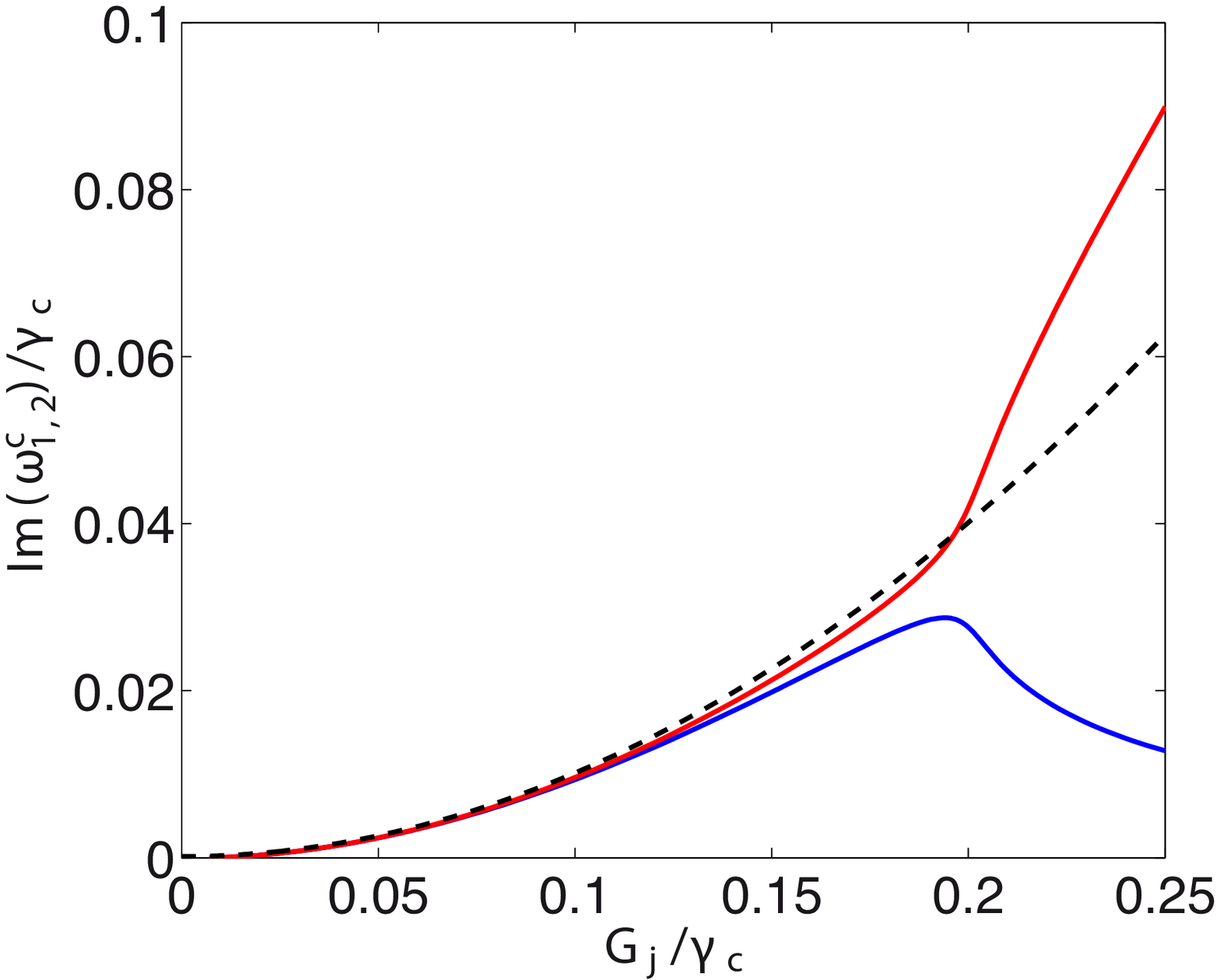}
    \caption{Values of $\omega_1^c$, $\omega_2^c$ in the weak-coupling
      regime. Left: real part of the frequencies and right: effective
      damping. On the left, the dashed lines show the approximation
      with a frequency independent coupling $G_{12}$ and
      disregarding the optical spring effect
      (Eq.~(\ref{eq:opticalspring})), whereas on the right, the dashed
      line shows the imaginary part of the effective frequencies
      $\omega_j^e$. For the damping, the coupling between the
      resonators shows up only in the strong-coupling regime. Here and in Figs.~\ref{fig:q_out}-\ref{fig:exact} the
      curves have been calculated with $\Delta=5 \gamma_c$,
      $\omega_{1,2}=|\Delta| \pm 0.0075\Delta$ and
      $\gamma_1=\gamma_2=3\times 10^{-5}|\Delta|$, close to the
      experimental values, and for simplicity assuming $G_1=G_2$.}
  \label{fig:cfreqs}
\end{figure}

\begin{figure}%[!ht]
  \includegraphics[width=0.45\textwidth]{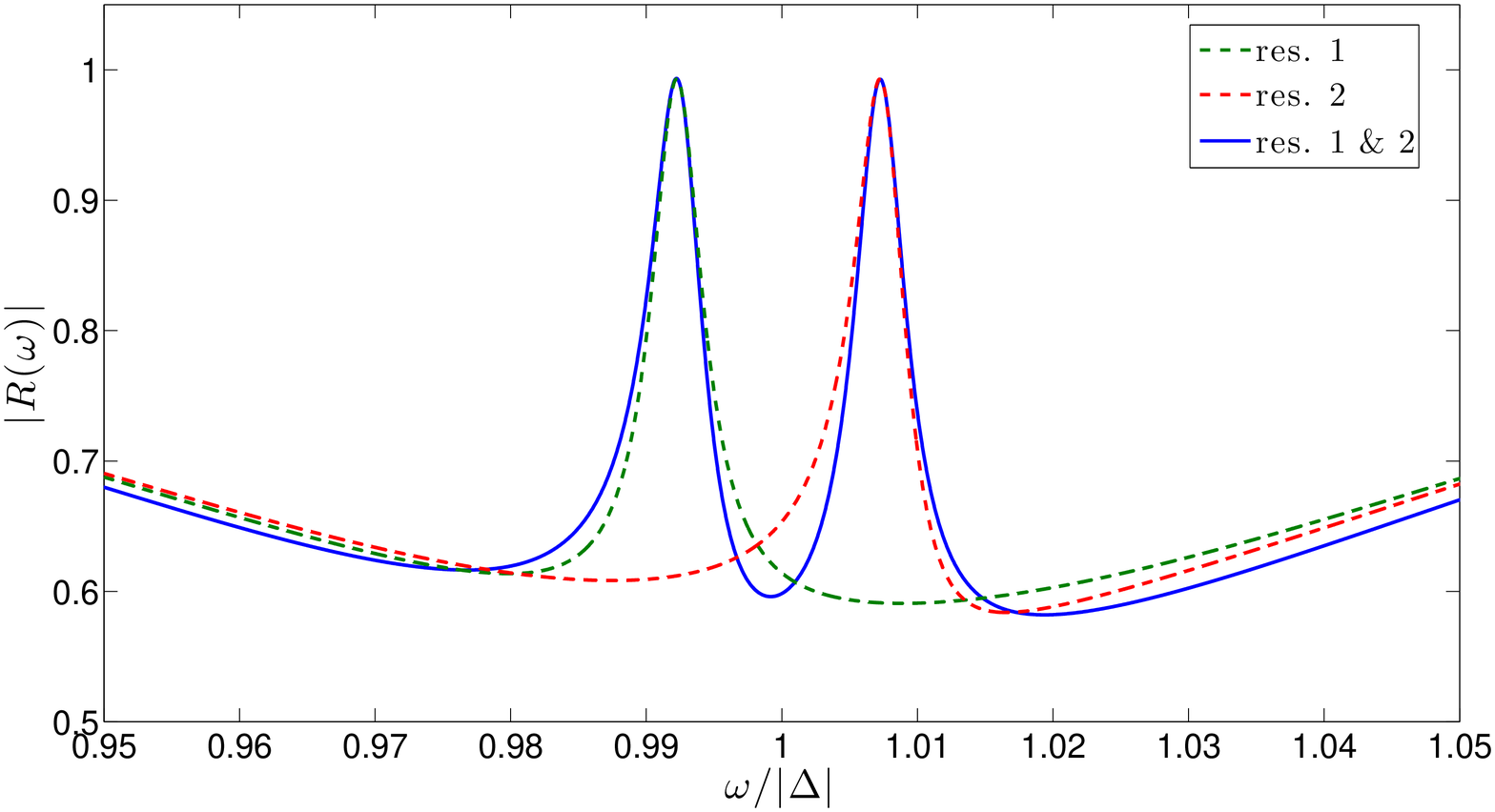}
    \includegraphics[width=0.45\textwidth]{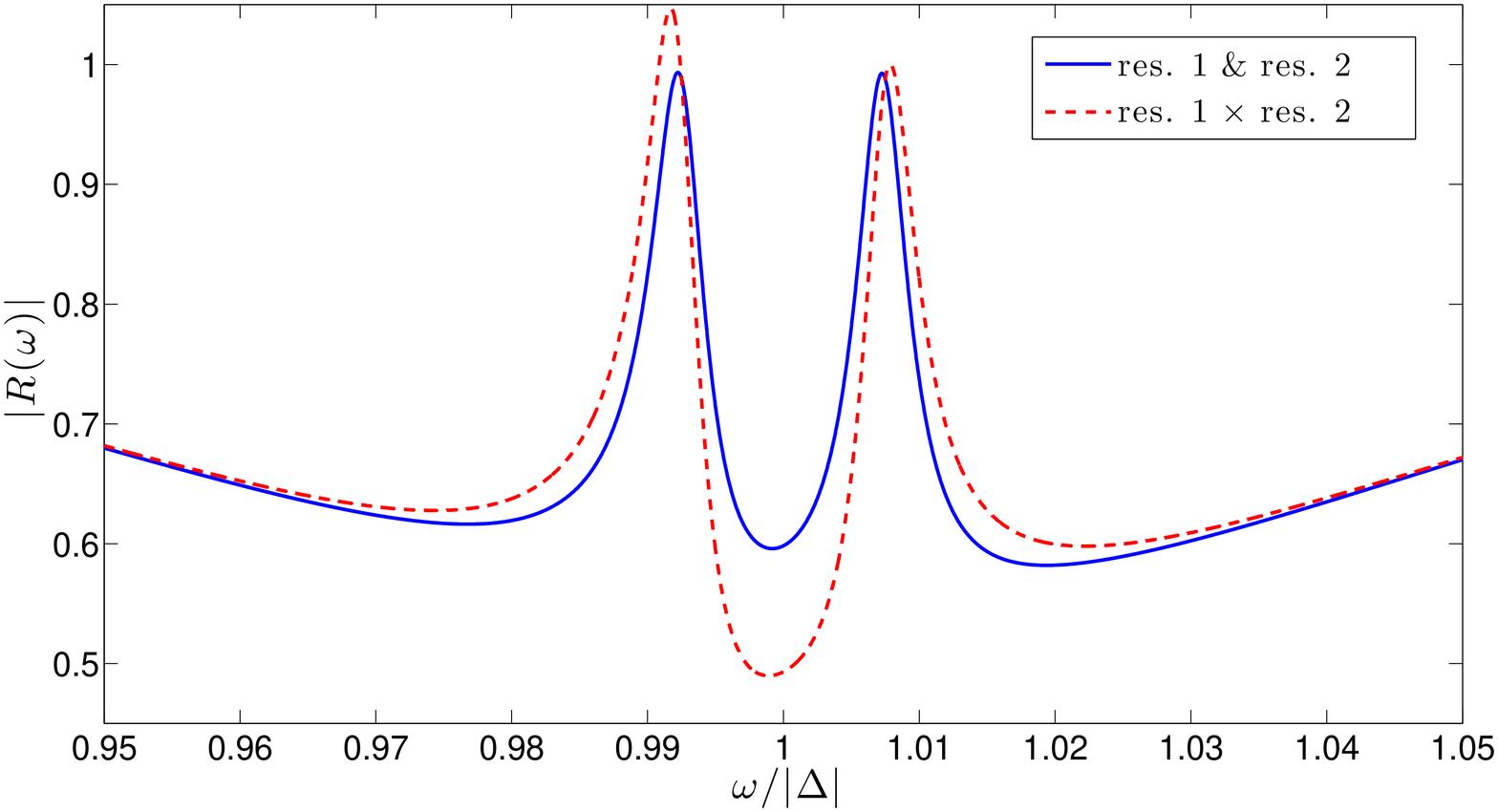}
    \caption{(left)Comparison of the reflection probability
      $|R(\omega)|$ for a single oscillator vs. that of two
      oscillators. (right) Comparison of the reflection probabilities
      when including (solid line, calculated with $\omega_j^c$) or
      excluding (dashed line, with $\omega_j^e$) the coupling induced
      normalization of the frequencies \vari{($G=0.1 \gamma_c$).}}
  \label{fig:comparison}
\end{figure}

\begin{figure}%[!ht]
    \includegraphics[width=0.5\textwidth]{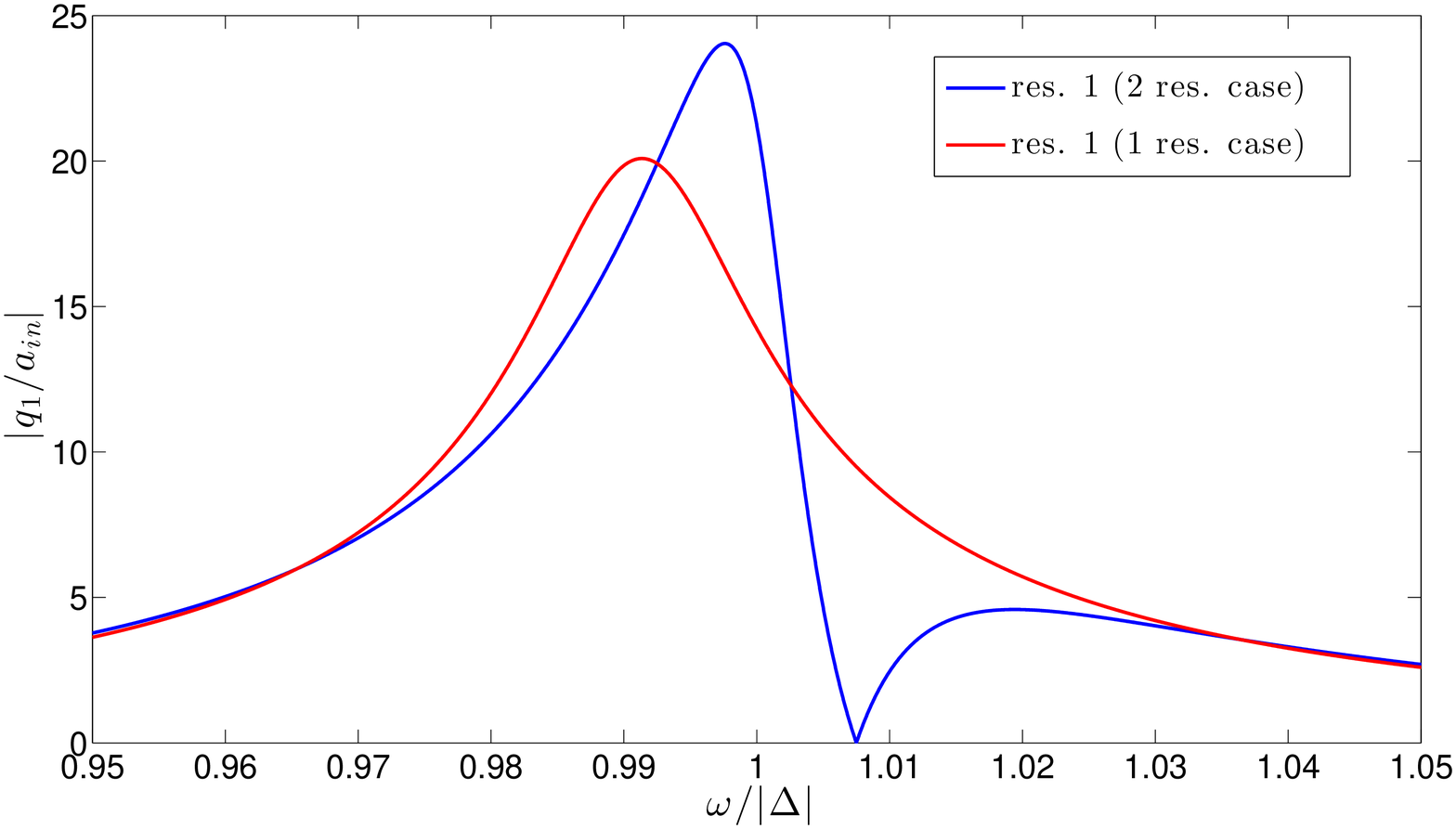}
    \caption{Mechanical spectrum of $|q_1|$ in the case of two (blue
      line) and one (red line) mechanical resonators, showing the frequency
      shift due to the coupling with the other resonator, an analogous
    situation would be observed for $|q_2|$ ($G=0.1 \gamma_c$).}
  \label{fig:q_out}
\end{figure}

\subsection{Strong-coupling regime}
\label{sec:strongcouplign}

In the strong-coupling limit $\tilde \chi G_1^2 G_2^2 \gtrsim
(\omega_1^e-\omega_2^e)^2$, $\omega_{1,2}^c$ can be expressed as
\begin{equation}
  \label{eq:22}
  \omega_{1,2}^c=\frac{1}{2}\left[\omega_1+\omega_2-
                           \tilde{\chi}\left(\ G_1^2+
                            G_2^2\right)/2 \pm \sqrt{\tilde{\chi}^2\left(G_1^2+G_2^2\right)^2/4}\right],
\end{equation}
from which we obtain two equations for $\omega_1^c$ and $\omega_2^c$
\begin{align}
  \label{eq:23}
  \omega_1^c&= \frac{\omega_1+\omega_2}{2}+\tilde{\chi}(\omega_1^c)\left(G_1^2+ G_2^2\right)/2 \\
  \omega_2^c&= \frac{\omega_1+\omega_2}{2}.
\end{align}
The second equation defines an asymptotic {\it dark mode} residing at
the frequency between the two mechanical frequencies, and having a
damping $(\gamma_1+\gamma_2)/2$, i.e., typically much lower than that
of the other dressed modes. For the first equation we have to include
the full frequency dependence of $\tilde \chi(\omega)$ from
Eq.~\eqref{eq:apprrespfuncs}. Multiplying the equation by $\tilde
\chi^{(-1)}$ then yields another second-order equation for
$\omega_1^c$, with the roots
\begin{equation}
\omega_1^c=\frac{1}{4} \left(-2 \Delta +\omega _1+\omega
   _2+i \gamma _c\pm \sqrt{\left(-i \gamma _c+2 \Delta +\omega _1+\omega _2\right){}^2+8 G_1^2+8 G_2^2}\right).
\label{eq:strongcouplingomega}
\end{equation}
For simplicity, we consider in the following the case
$\Delta=-(\omega_1+\omega_2)/2$ which simplifies the expression to
\begin{equation}
  \omega_1^c=-\Delta + \frac{1}{4}\left(i\gamma_c\pm \sqrt{8 (G_1^2+G_2^2)-\gamma_c^2}\right).
\end{equation}
When $G_1^2+G_2^2 > \gamma_c^2/8$, the real part of the
frequencies tends to
\begin{equation}
\omega_1^c \rightarrow -\Delta \pm G,
\end{equation}
where $G=\sqrt{(G_1^2+G_2^2)/2}$. Moreover, the linewidth of
these modes, whenever $G > \gamma_c/4$, is given by $\gamma_c/2$. In
this case these frequencies and linewidths can be seen in the cavity
response, as the cavity response function $|R(\omega)|$ has three
dips: the outer dips, of width $\sim \gamma_c/2$ correspond to the
(bright) modes $\omega_1^c$, whereas the inner narrow dip corresponds
to the (dark) mode $\omega_2^c$ (see Fig.~2e of the main text).

% While the second equation unambiguously defines , in order to get
% a simple expression for $\omega_1^c$, we have to simplify the
% frequency dependence of $\tilde{\chi}$. Far enough from resonance
% (where the approximation leading to eq. \eqref{eq:22} is valid),
% $\tilde{\chi}(\omega_1^c)$ can be approximated as
% \begin{equation}
%   \label{eq:24}
%   \tilde{\chi}(\omega_1^c) \simeq \frac{\Delta
%     (\Delta+\omega_1^c)(\Delta-\omega_1^c)}{
%     (\Delta+\omega_1^c)^2(\Delta-\omega_1^c)^2} \simeq \frac{1}{2\left(\omega_1^c-\Delta\right)}
% \end{equation}
% where, in the last equality, we have  assumed that $\omega_1^c - \Delta
% \simeq -2 \Delta$. 
% If we consider a situation where $-\Delta=\frac{\omega_1+\omega_2}{2}$,
% we can write, substituting eq. \eqref{eq:24} into eq. \eqref{eq:23}
% \begin{equation}
%   \label{eq:26}
%    \omega_1^c=-\Delta  + \frac{G}{4\left(\omega_1^c-\Delta\right)}
% \end{equation}
% where $G=\sqrt{G_1^2+G_2^2}$.
% The expression given by eq. \eqref{eq:26}, allows us to re-express the product
%  $(\omega-\omega_1^c)(\omega=\Delta)$ as 
% \begin{equation}
%   \label{eq:6}
%   (\omega-\omega_{1+}^c)(\omega-\omega_{1-}^c)
% \end{equation}
% with 
% \begin{equation}
%   \label{eq:25}
%  \omega_{1\pm}^c=\Delta \pm G
% \end{equation}

Let us try to understand these modes from starting the equations of
motion by redefining the mechanical motion to relative and center of
mass motion (weighted by the couplings $G_j$), i.e.,
\begin{align}
  \label{eq:17}
 q_a=\left(  G_2 q_1 -  G_1 q_2 \right)/\left(
    G_2 + G_1 \right) \\ 
 q_s=\left(  G_1 q_1 +  G_2 q_2 \right)/\left(
    G_2 + G_1 \right).
\end{align}
Below, we show that in the strong-coupling limit the above frequency
$\omega_2^c$ describes the ``dark'' mode $q_a$ and the two frequencies
$\omega_{1\pm}^c$ the ``bright'' modes $q_s$.

The equations of motion for these modes are
\begin{subequations}
\begin{align}
  \dot{a}&=i \Delta a + i(G_1+G_2) q_s + \sqrt{\gamma_E}a_{\rm in}\\
  \dot{q}_s&=\omega_\Sigma p_s + \omega_\Delta p_a \\
  \dot{q}_a&=\omega_\Sigma p_a + \omega_\Delta p_s  \label{eq:19b}\\
  \dot{p}_s&=-\omega_\Sigma q_s + \omega_\Delta q_a - \gamma p_s +
  \frac{ G_1  G_2}{ G_1+ G_2} \left( a^\dagger + a\right) \\
  \dot{p}_a&=-\omega_\Sigma q_a + \omega_\Delta q_s - \gamma p_a
\end{align}
\label{eq:commonq}
\end{subequations}
with $\omega_\Sigma=\left(\omega_1+\omega_2\right)/2$ and
$\omega_\Delta=\left(\omega_1-\omega_2\right)/2$. We have assumed that
the linewidths of the two mechanical resonators are the same. If there
is no frequency mismatch, the symmetric and antisymmetric modes are
uncoupled.

Solving Eqs.~\eqref{eq:commonq}, or substituting Eqs.~(\ref{eq:q1},\ref{eq:q2})
into \eqref{eq:17} (and using the complex frequency notation) 
% yields
% \begin{align}
%   \label{eq:13}
%   q_1=\frac{ G_1}{\left(
%     G_2 + G_1 \right)\sqrt{2 \gamma}} 
%          \frac{\omega-\omega_2}{\omega-\omega_2^c} 
%          \frac{1}{\omega-\omega_1^c} \frac{1}{\omega+\Delta} a_{\rm in} \\
%   q_2=\frac{ G_2}{\left(
%     G_2 + G_1 \right)\sqrt{2 \gamma}} 
%          \frac{\omega-\omega_1}{\omega-\omega_1^c} 
%          \frac{1}{\omega-\omega_2^c} \frac{1}{\omega+\Delta} a_{\rm in}
% \end{align}
we obtain, in the strong-coupling limit
\begin{align}
  \label{eq:20}
  q_s&=\frac{4  G_1  G_2 }{\left(
    G_2 + G_1 \right)}  \frac{\omega-\omega_\Sigma}
         {\left(\omega-\omega_{1+}^c\right)\left(\omega-\omega_{1-}^c\right)\left(\omega-\omega_{2}^c\right)}
  \sqrt{\gamma_E} a_{\rm in} \\
  q_a&=\frac{4  G_1  G_2}{\left(
    G_2 + G_1 \right)}\frac{\omega_\Delta}
     {\left(\omega-\omega_{1+}^c\right)\left(\omega-\omega_{1-}^c\right)\left(\omega-\omega_{2}^c\right)} \sqrt{\gamma_E}a_{\rm in} .
\end{align}
Considering that $\omega_2^c = \omega_\Sigma$, and assuming for simplicity
$ G_1 =  G_2 \equiv  G$, we
have, when $\omega \simeq \omega_{1\pm}^c $,
%%% previous equations miss \sqrt{\gamma_E}
\begin{align}
  \label{eq:20b}
  q_{s+}&=\frac{\sqrt{\gamma_E} }{
    (\omega-\omega_{1+}^c)} a_{\rm in}\\
  q_{s-}&=-\frac{\sqrt{\gamma_E} }{
    (\omega-\omega_{1+}^c)} a_{\rm in}
\end{align}
the cavity field becomes
\begin{align}
  \label{eq:27}
  X=\frac{\sqrt{ \gamma_E}}{2(\omega-\omega_{1\pm}^c)} a_{\rm in}.
\end{align}
% Furthermore, we can express the cavity field as a function of
% $q_{s\pm}$, obtaining
% \begin{equation}
%   \label{eq:28}
%   a= \sqrt{2 \gamma} q_{s\pm}.
% \end{equation}
Analogously, when $\omega \simeq -\Delta$, we get for $q_a$
\begin{equation}
  \label{eq:20c}
  q_a=\frac{2\omega_\Delta}{G \left(\omega-\omega_2^c\right)}\sqrt{\gamma_E} a_{\rm in}. 
\end{equation}
In this case the cavity field around the dark mode resonance frequency becomes
\begin{equation}
  \label{eq:29}
  X=-\frac{\left(\omega_1+\Delta\right)\left(\omega_2+\Delta\right)}{G^2\left(\Delta-\omega_2^c\right)}\sqrt{\gamma_E}a_{\rm in}.
\end{equation}
% or, if expressed in terms of $q_a$
% \begin{equation}
%   \label{eq:15}
%   a=-\sqrt{\frac{\gamma}{2}} \frac{\omega_\Delta}{\sqrt{2} G} q_a
% \end{equation}
% where we have assumed that $\omega_1=\Delta+\delta\omega$ and
% $\omega_2=\Delta-\delta\omega$.

From Eq. \eqref{eq:20c} we see that, in the strong coupling limit,
the mechanical mode $q_a$ is asymptotically decoupled from the input
as well as from the cavity field, allowing to identify this mode as a
dark mode.

From a somewhat different perspective, the above picture and the
values of the effective frequencies can be recovered by diagonalizing
the equations of motion for $q_a$, $p_a$, $q_s$, $p_s$, and $a$.
\begin{figure}%[!ht]
    \includegraphics[width=0.45\textwidth]{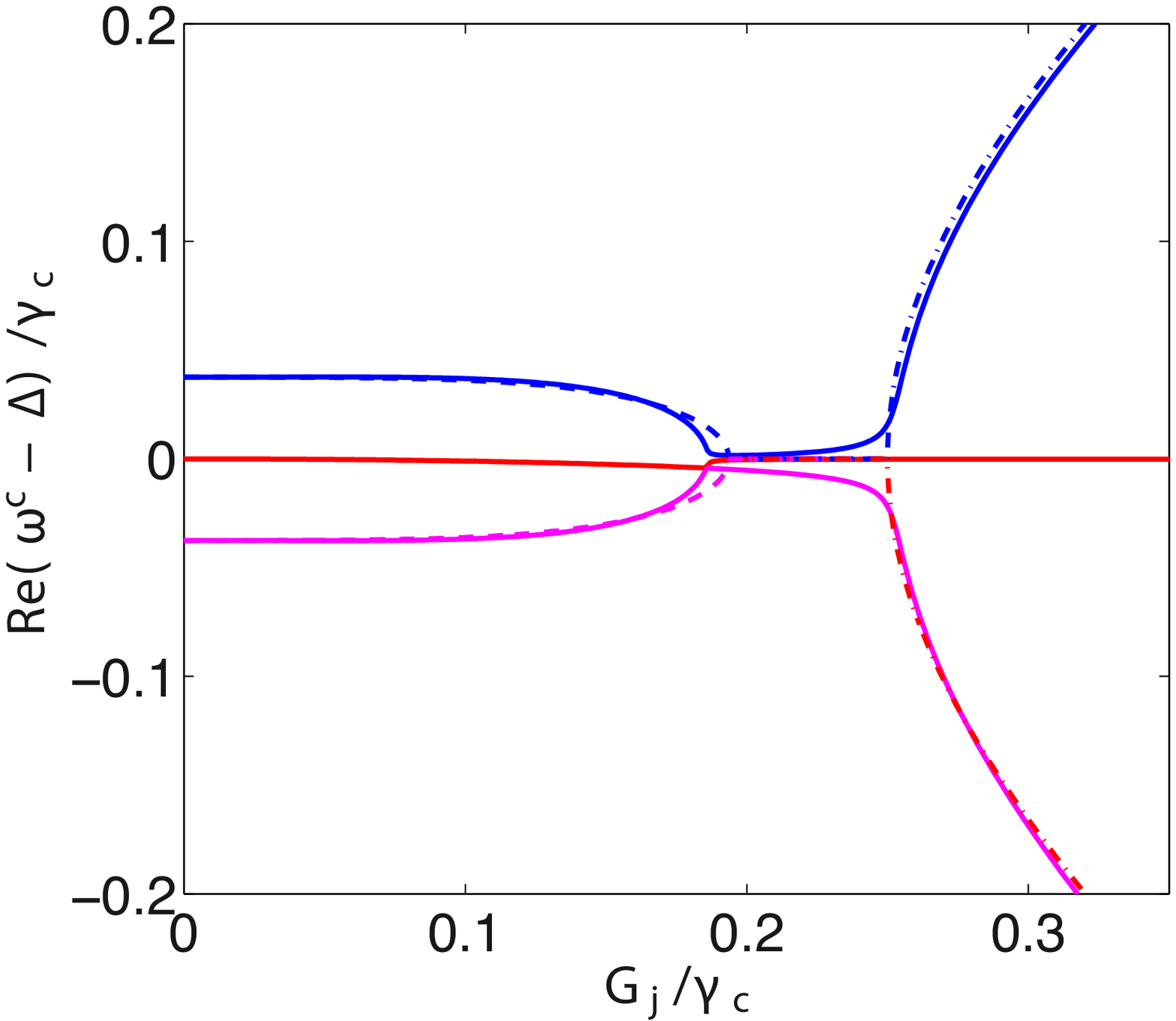}
    \includegraphics[width=0.45\textwidth]{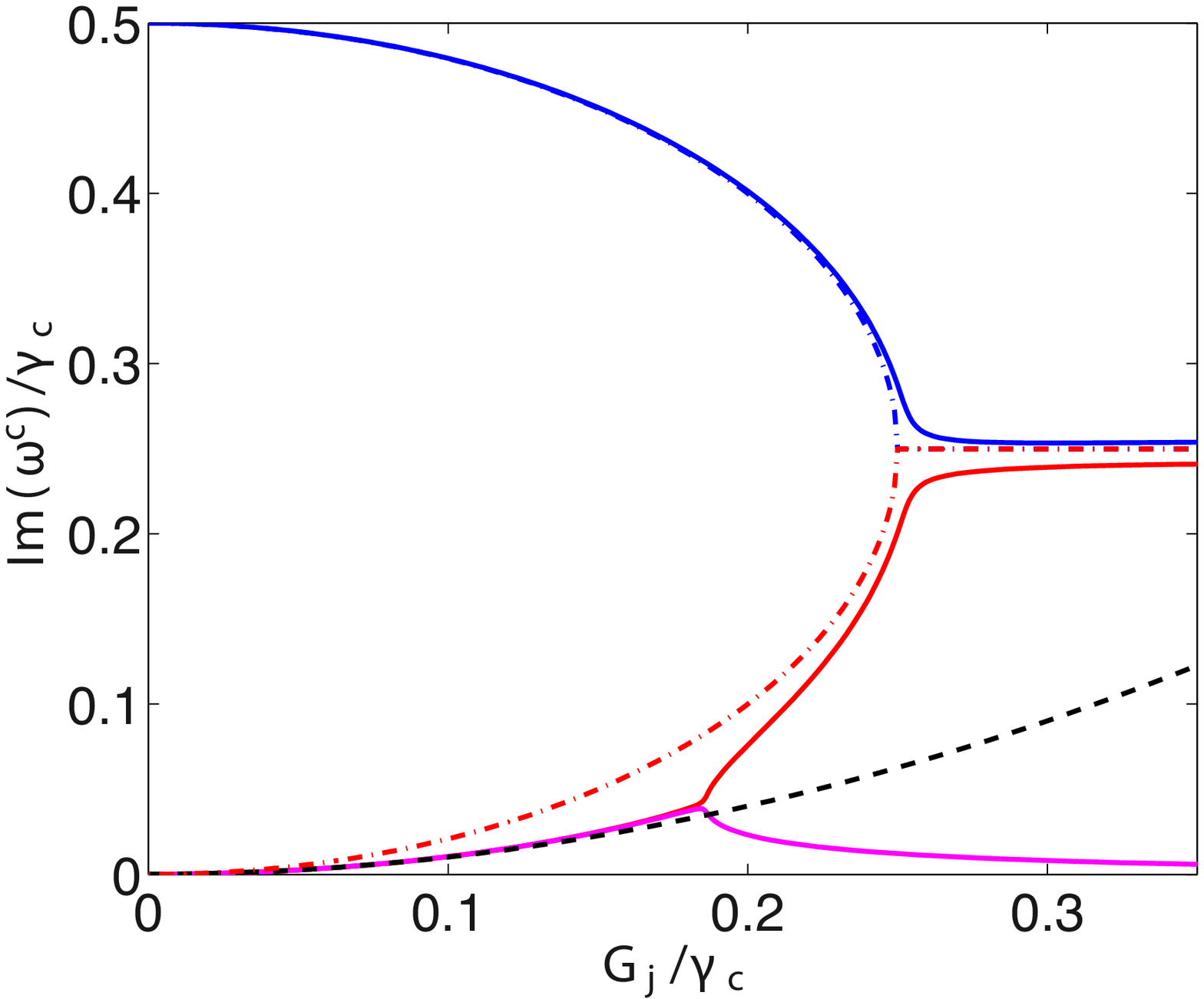}
    \caption{Comparison between the calculated values of
      $\omega_{1,2}^c$ with the values obtained by diagonalization of
      the equations of motion (solid lines). (Left:) Real part of the
      frequency. Dashed line is the weak-coupling expression obtained
      from Eq.~\eqref{eq:12} by using Eq.~\eqref{eq:gmech} and the
      dash-dotted line is obtained from the strong coupling expression
      Eq.~\eqref{eq:strongcouplingomega}. (Right): Effective
      linewidths of the modes. Black dashed line is from
      eqs.~(\ref{eq:opticalspring}), showing the regular optically
      induced damping, whereas the dash-dotted lines are
      obtained from Eq.~\eqref{eq:strongcouplingomega}. These
      frequencies have been calculated with the same parameters as in
      the previous figures.}.
  \label{fig:exact}
\end{figure}
In Fig.\ref{fig:exact}, we have compared the relevant
eigenfrequencies around $-\Delta$ as obtained by the diagonalization of
Eqs.~\eqref{eq:commonq} with the strong and weak coupling expansion
for $\omega_{1 \pm}^c$ and $\omega_2^c$. In the weak-coupling regime
the eigenstates correspond (in the limit $ G_{1,2}\to 0$) to the
normal modes of the uncoupled system (i.e. cavity, mechanical
resonator 1, mechanical resonator 2, cavity field). The corresponding
eigenfrequencies match the expression given in Eq. \eqref{eq:12} in
the case where $\tilde{\chi}$ is independent of $\omega$.

In the strong-coupling regime these modes correspond to the coherent
superposition of the cavity field and $q_{s\pm}$ modes or zero cavity
field and $q_a$ modes. We can gain further insight about these modes
by writing the strong-coupling version ($\omega_\Delta \ll  G$) of the
equation of motion \eqref{eq:19b} in the sideband-resolved regime
\begin{align}
  \label{eq:16}
   &\dot{a}_a= -i \omega_\Sigma a_a -\frac{\gamma}{2} a_a \\
   &\dot{a}_s= -i \omega_\Sigma a_s -\frac{\gamma}{2} a_s + \frac{i
      G}{\sqrt{2}} a\\ 
   &\dot{a} = i \Delta a + i  \sqrt{2}G a_s - \frac{\gamma_c}{2} a.
\end{align}
Its eigenfrequencies are given by 
\begin{align}
  \label{eq:33}
  &\omega_A=\omega_\Sigma \\
  &\omega_{B\pm}=-\Delta \pm 1/4 \sqrt{16 G^2-\gamma_c^2}.
\end{align}
Moreover, the eigenvector associated to $\omega_A$, corresponds
(asymptotically) to a vector for which $a$ and $a_s$ are zero, while
the eigenvectors associated to $\omega_{B\pm}$ correspond to vectors
for which $a_a$ is zero, allowing thus to identify $\omega_A$ with the
frequency of the dark mode and $\omega_{B\pm}$ to that of the bright
modes.

The coupling $G=\gamma_c/4$ corresponds to the onset of the
strong-coupling regime where, asymptotically, the eigenmodes
correspond to the dark and bright modes. Moreover the limit $ G \gg
\gamma_c$ yields
\begin{equation}
  \label{eq:34}
  \omega_{B\pm}=-\Delta \pm G,  
\end{equation}
allowing us to identify the symmetric mode $q_s$ with the mechanical
component of the bright mode, as it can be seen from the expression
giving the strong-coupling expansion of $\omega_1^c$ (see
eqs.~(\ref{eq:12})).

%\end{multicols}long equation goes here
\end{widetext}

\end{document}